\documentclass[a4paper,fleqn,usenatbib]{mnras}

\usepackage{graphicx}	% Including figure files
\usepackage{amsmath}	% Advanced maths commands
\usepackage{amssymb}	% Extra maths symbols
\usepackage{pdflscape}
\newcommand\mnrassub{MNRAS, in press}

\title[Physical reality of open clusters]{On the physical reality of overlooked open clusters}

\author[A.E. Piatti]{
Andr\'es E. Piatti$^{1,2}$\thanks{E-mail: andres@oac.unc.edu.ar}\\
$^{1}$Consejo Nacional de Investigaciones Cient\'{\i}ficas y T\'ecnicas, Av. Rivadavia 1917, 
C1033AAJ, Buenos Aires, Argentina\\
$^{2}$Observatorio Astron\'omico, Universidad Nacional de C\'ordoba, Laprida 854, 5000, 
C\'ordoba, Argentina\\
}

\date{Accepted XXX. Received YYY; in original form ZZZ}

\pubyear{2016}

\begin{document}
\label{firstpage}
\pagerange{\pageref{firstpage}--\pageref{lastpage}}
\maketitle

\begin{abstract}
We present  $UBVRI$ and $CT_1T_2$ photometry for fifteen catalogued open clusters
of relative high brightness and compact appearance. From these unprecedented
photometric data sets, covering wavelengths from the blue up to the near-infrared,
we performed a thorough assessment of their reality as stellar aggregates.
We statistically assigned
to each observed star within the object region a probability of being a fiducial feature
of that field in terms of its local luminosity function, colour distribution and 
stellar density. Likewise, we used accurate parallaxes
and proper motions measured by the {\it Gaia} satellite to help our
decision on the open cluster reality. Ten catalogued aggregates did not show any 
hint of being real physical systems; three of them had been assumed to
be open clusters in previous studies, though. On the other hand, we estimated reliable fundamental
parameters for the remaining five studied objects, which were confirmed as real
open clusters. They resulted to be clusters distributed in a wide age range, 
8.0 $\le$ $\log$($t$ yr$^{-1}$) $\le$ 9.4, of solar metal content and placed between 2.0 and 5.5 kpc
from the Sun. Their ages and metallicities are in agreement with the presently known
picture of the spatial distribution of open clusters in the Galactic disc.
\end{abstract}


%%%%%%%%%%%%%%%%%%%%%%%%%%%%%%%%%%%%%%%%%%%%%%%%%%

%%%%%%%%%%%%%%%%% BODY OF PAPER %%%%%%%%%%%%%%%%%%

\section{Introduction}

According to the most updated version of the open cluster catalogue compiled by 
\citet[][version 3.4 as of 2016 January]{detal02}, a very limited number of objects 
have been studied with some detail. Indeed, many of the catalogue's entries correspond
to stellar overdensities \citep[e.g.][]{collinder1931,ruprecht1966} 
which deserve further analyses in order to confirm their physical nature. 
Hence, we have recently made use of available multi-band imaging with the aim of
improving the statistics of well studied open clusters. We assessed the reality of 
the observed catalogued objects as real open clusters and, 
for those confirmed, estimated their structural and fundamental parameters, studied 
their dynamical evolution, etc \citep[e.g.][]{piatti16b}.
 
In this paper we complete the analysis of every open cluster observed through
the Johnson $UBV$, Kron-Counsins $RI$ and Washington $CT_1T_2$ photometric passbands during the observing
campaign carried out at the Cerro Tololo Inter-American Observatory (CTIO), Chile, in 2011
January 31--February 4  (CTIO programme \#2011A-0114, PI: Clari\'a). The objects had been
originally selected to examine how the
abundance gradient in the Galactic disc evolved in time and along different Galactic longitudes 
by comparing the abundance gradients corresponding to various groups of open clusters of 
different ages and positions. The greater the number of open clusters with well determined parameters, 
the more precise and detailed the analysis of the metal abundance gradient in the Galactic disc 
as well as its evolution over time. 

The present sample comprises unstudied or poorly studied catalogued open clusters that are 
relatively bright, are composed by a small number of stars and have a relatively compact appearance.
In Section 2 we describe the collection and reduction of the available photometric data  
in order to build extensive and reliable data sets. We carefully evaluated the physical reality
of the cluster sample from colour-magnitude and colour-colour diagrams in Section 3,
while in Section 4 we estimate the fundamental parameters of the confirmed stellar
aggregates and discuss the results. Finally, Section 7 summarizes the main conclusions of
 this work.

\section{Data collection and reduction}

The National Optical Astronomy Observatory 
(NOAO) Science Data Management (SDM) Archives\footnote{http://www.noao.edu/sdm/archives.php.}
interface were used to download the whole set of  publicly available
calibration and programme 
images . 
 The data processing, the calibration to the
standard photometric systems and the point spread function photometry for the programme fields were
carried out altogether with some peviously studied objects, as described in \citet{piattietal17}.
We included here two catalogued open clusters (ESO\,371-25 and 561-5) for which
Washington $CT_1$ images are publicly available and had not paid particular attention to them until now.
The data reduction and photometry for these two objects were performed as described in 
\citet{piattietal2011}.
The log of the observations for the presently studied objects with the main astrometric and 
observational information is summarized in Table~\ref{tab:table1}.

\begin{table*}
\caption{Observations log of selected star clusters.}
\label{tab:table1}
\begin{tabular}{@{}lccccccc}\hline
Cluster  &R.A.      &Dec.     &{\it l} &b       &  filter & exposure & airmass\\
         &(h m s)   &($\degr$ $\arcmin$ $\arcsec$)&(\degr)&(\degr)&  &  (sec)  & \\
\hline

Alessi\,14   & 6 30 25.9&+09 44 28  & 201.8790 &  -0.1796 & $U$ & 90, 480& 1.35, 1.35  \\
             &          &           &          &          & $B$ & 60, 300& 1.32, 1.32    \\
             &          &           &          &          & $V$ & 40, 180& 1.31, 1.31    \\
             &          &           &          &          & $R$ & 20, 120 & 1.31, 1.31 \\
             &          &           &          &          & $I$ & 10, 90& 1.30, 1.30 \\
             &          &           &          &          & $C$ & 90, 420 & 1.33, 1.33   \\
             &          &           &          &          &  &   &   \\
ESO\,211-9   & 9 16 44.2&-50 16 59  & 271.9430 & -0.7949  & $U$ & 150, 600 & 1.07, 1.07    \\
             &          &           &          &          & $B$ &100, 400 & 1.09, 1.09    \\
             &          &           &          &          & $V$ & 90, 200 &  1.10, 1.10  \\
             &          &           &          &          & $R$ & 30, 140 & 1.11, 1.11   \\
             &          &           &          &          & $I$ & 20, 120  & 1.12, 1.12   \\
             &          &           &          &          & $C$ & 120, 580  & 1.08, 1.08    \\
             &          &           &          &          &  &   &   \\
ESO\,260-6   & 8 46 17.3&-47 41 06  & 266.6933 & -2.8282  & $U$ &60, 480 & 1.06, 1.06   \\
             &          &           &          &          & $B$ & 60, 120 & 1.05, 1.05    \\
             &          &           &          &          & $V$ &20, 60, 180 & 1.05, 1.05, 1.05   \\
             &          &           &          &          & $R$ & 20, 120  & 1.05, 1.05   \\
             &          &           &          &          & $I$ &10, 90& 1.05, 1.05    \\
             &          &           &          &          & $C$ & 50, 420  & 1.05, 1.05    \\
             &          &           &          &          &  &   &   \\
ESO\,315-14  & 9 35 25.1&-39 32 03  & 266.7956 &  9.2082  & $U$ &90, 480 & 1.04, 1.04    \\
             &          &           &          &          & $B$ & 60, 320 & 1.07, 1.07   \\
             &          &           &          &          & $V$ & 60, 180 & 1.09, 1.10    \\
             &          &           &          &          & $R$ & 30, 120 & 1.11, 1.11   \\
             &          &           &          &          & $I$ &15, 90 & 1.12, 1.12   \\
             &          &           &          &          & $C$ &80, 400 & 1.05, 1.06   \\
             &          &           &          &          &  &   &   \\
ESO\,371-25  & 8 53 00.0&-35 28 01  & 258.0026 & 5.8612   & $C$ & 300, 300 & 1.01, 1.01   \\
             &          &           &          &          & $R$ &  10, 10, 30, 30& 1.01, 1.01, 1.01, 1.01  \\
             &          &           &          &          &  &   &   \\
ESO\,383-10  &13 31 30.0&-35 03 56  & 312.1324 &27.0895   & $U$ &15, 180 & 1.02, 1.02    \\
             &          &           &          &          & $B$ &10, 120 & 1.01, 1.01   \\
             &          &           &          &          & $V$ &7, 80 & 1.01, 1.01    \\
             &          &           &          &          & $R$ &5, 60  & 1.01, 1.01   \\
             &          &           &          &          & $I$ &4, 60 & 1.01, 1.01   \\
             &          &           &          &          & $C$ &15, 180 & 1.02, 1.02    \\
             &          &           &          &          &  &   &   \\
ESO\,430-9   & 8 02 24.0&-29 46 01  & 247.1599 & 0.5167   & $U$ & 90, 450  & 1.00, 1.00   \\
             &          &           &          &          & $B$ & 60, 200 & 1.02, 1.01    \\
             &          &           &          &          & $V$ & 60, 200 & 1.02, 1.02   \\
             &          &           &          &          & $R$ & 15, 120 & 1.03, 1.03    \\
             &          &           &          &          & $I$ & 10, 10, 90 & 1.04, 1.04, 1.03    \\
             &          &           &          &          & $C$ &80, 400 & 1.01, 1.01    \\
             &          &           &          &          &  &   &   \\
ESO\,437-61  &10 48 03.0&-29 23 30  & 273.0624 &26.2286   & $U$ & 90, 400 & 1.07, 1.07    \\
             &          &           &          &          & $B$ &60, 300 & 1.04, 1.04    \\
             &          &           &          &          & $V$ &40, 200& 1.03, 1.03   \\
             &          &           &          &          & $R$ &20, 120 & 1.02, 1.03 \\
             &          &           &          &          & $I$ &10, 90 & 1.02, 1.02, 1.02   \\
             &          &           &          &          & $C$ &80, 360 & 1.05, 1.06   \\
             &          &           &          &          &  &   &   \\
ESO\,493-3   & 7 39 42.0&-27 17 00  & 242.4794 &-2.5035   & $U$ & 60, 240& 1.08, 1.09 \\
             &          &           &          &          & $B$ &20, 150& 1.06, 1.06 \\
             &          &           &          &          & $V$ &   10, 120& 1.05, 1.05 \\
             &          &           &          &          & $R$ &  10, 90 & 1.03, 1.04 \\
             &          &           &          &          & $I$ & 10, 60 & 1.03, 1.03 \\
             &          &           &          &          & $C$ &   60, 180& 1.07, 1.07 \\
             &          &           &          &          &  &   &   \\
ESO\, 561-5  & 7 59 18.0&-22 41 00  & 240.7668 &+3.6558   & $C$ & 300 & 1.08 \\
             &          &           &          &          & $R$ & 30 & 1.08  \\

\hline
\end{tabular}
\end{table*}

\setcounter{table}{0}
\begin{table*}
\caption{continued.}
\label{tab:table1}
\begin{tabular}{@{}lccccccc}\hline
Cluster  &R.A.      &Dec.     &{\it l} &b       &  filter & exposure & airmass\\
         &(h m s)   &($\degr$ $\arcmin$ $\arcsec$)&(\degr)&(\degr)&  &  (sec)  & \\
\hline

Hogg\,9      &10 58 24.5&-59 03 11  & 288.8399 &+0.6910   & $U$ &80, 480 &1.16, 1.17    \\
             &          &           &          &          & $B$ & 60, 300 & 1.19, 1.19   \\
             &          &           &          &          & $V$ & 40, 180 & 1.20, 1.20   \\
             &          &           &          &          & $R$ &10, 20, 100 & 1.22, 1.21, 1.21   \\
             &          &           &          &          & $I$ & 10, 80 & 1.22, 1.20    \\
             &          &           &          &          & $C$ &70, 420   & 1.18, 1.18   \\
             &          &           &          &          &  &   &   \\
Miller\,1    &9 25 40.8 &-53 13 30  & 275.0021 &-1.8977   & $U$ &90, 500 &1.09, 1.09     \\
             &          &           &          &          & $B$ &80, 320 & 1.08, 1.09    \\
             &          &           &          &          & $V$ &60, 180 & 1.09, 1.09   \\
             &          &           &          &          & $R$ &20, 120 &1.10, 1.10    \\
             &          &           &          &          & $I$ &15, 90 & 1.10, 1.10   \\
             &          &           &          &          & $C$ &90, 420 & 1.09, 1.09   \\
            &          &           &          &          &  &   &   \\
NGC\,5269    &13 44 44.0&-62 54 54  & 308.9554 &-0.6682   & $U$ & 90, 540 & 1.21, 1.21   \\
             &          &           &          &          & $B$ & 80, 80, 300   & 1.22, 1.22, 1.21   \\
             &          &           &          &          & $V$ &60, 200 & 1.22, 1.22   \\
             &          &           &          &          & $R$ & 15, 120  & 1.23, 1.22    \\
             &          &           &          &          & $I$ &15, 90 & 1.23, 1.23    \\
             &          &           &          &          & $C$ &80, 420 & 1.21, 1.20    \\
             &          &           &          &          &  &   &   \\
Ruprecht\,5  &6 55 14.2 &-18 36 00  & 230.0043 &-7.5541   & $U$ & 80, 540  & 1.05, 1.05   \\
             &          &           &          &          & $B$ &60, 360 & 1.09, 1.08   \\
             &          &           &          &          & $V$ &60, 220 & 1.11, 1.11   \\
             &          &           &          &          & $R$ &50, 140 & 1.11, 1.11   \\
             &          &           &          &          & $I$ &10, 20, 100 & 1.13, 1.12, 1.12   \\
             &          &           &          &          & $C$ &80, 480 & 1.07, 1.07   \\
             &          &           &          &          &  &   &   \\
Ruprecht\,15 &7 19 31.7 &-19 37 48   & 233.5384&-2.9061   & $U$ & 90, 480 & 1.02, 1.02   \\
             &          &           &          &          & $B$ & 60, 240  & 1.04, 1.04   \\
             &          &           &          &          & $V$ &40, 150 & 1.05, 1.05   \\
             &          &           &          &          & $R$ &20, 100 & 1.05, 1.06   \\
             &          &           &          &          & $I$ &10, 80  & 1.06, 1.06    \\
             &          &           &          &          & $C$ &80, 420  & 1.03, 1.03   \\

\hline
\end{tabular}
\end{table*}

The final information for each cluster
field consists of a running number per star, its $x$ and $y$ coordinates, the mean $V$ 
magnitude, its rms error and the number of measurements, the colours $U-B$, $B-V$,
$V-R$, $V-I$ with their respective rms errors and number of measurements, the $T_1$
magnitude with its error and number of measurements, and the $C-T_1$ and $T_1-T_2$
colours with their respective rms errors and number of measurements.
Table~\ref{tab:table2} gives this information for Alessi\,14. Only a portion 
of this table is shown here for guidance regarding its form and content. The whole content 
of Table~\ref{tab:table2}, as well as those for the remaining cluster fields
(Tables 3-16), are available in the online version of the journal.

\section{colour-magnitude diagram analysis}

We based our analysis of the selected objects on their colour-magnitude diagrams (CMDs) 
and colour-colour (CC) diagrams,
once we estimated the open cluster extents, extrated the resulting multi-band photometry
for every stars located within the object areas and thoroughly judged their
reality as stellar aggregates.

In order to estimate the cluster radii we built stellar density radial profiles from their
geometrical centres. Those centres were adopted from single Gaussian fits  performed to the
stellar density distributions along the $x$ and $y$ directions by
using the {\sc ngaussfit} routine in the {\sc stsdas/iraf} package. We repeated the
fits for projected stellar density distributions built from the number of stars
counted within intervals of  20 (5.8), 40 (11.6), 60 (17.4), 80 (23.2) and 100 (29.0) pixel (arcsec) wide, and finally we 
averaged the five different Gaussian centres resulting a typical standard deviation of 
$\pm$  50 pixels ($\pm$ 14.5$\arcsec$) in all cases.

We traced the radial density profiles from star counts previously performed within
boxes of 50 pixels (14.5 arcsec) a side distributed throughout the whole FOV,
instead of employing rings around the centre of each cluster.
The chosen box size allowed us to sample the stellar spatial distribution statistically.
We followed this method since it does not require a 
complete circle of radius $r$ within the observed FOV to estimate the mean stellar
density at that distance. With a stellar density profile that extends far away from 
the cluster centre -but not too far so as to risk losing the local field-star signature- 
it is possible to estimate the background level with 
high precision, which is particularly useful when dealing with clusters composed by
a small number of stars.
The number of stars per unit area at a given radius $r$ was calculated 
through the expression:

\begin{equation}
(n_{r+25} - n_{r-25})/(m_{r+25} - m_{r-25}),
\end{equation}

\noindent where $n_r$ and $m_r$ represent the number of stars and boxes, respectively,  
included in a circle of radius $r$. The resulting density profiles 
expressed as number of stars per arcsec$^2$ are shown in  Fig.~\ref{fig:fig1}.
In the figure, we represent the constructed and
background subtracted density profiles with open and filled circles, respectively.
The average and corresponding rms error of the background level at any distance to the 
cluster centres were counted by using every available star
count measurement at that distance. Then, the mean background and its error was 
calculated by averaging all these latter values, and indicated by solid and dotted
horizontal lines in the figure. The errorbars of the background subtracted density 
profiles include  the mean error of the background level. 

As for the cluster radius, we considered here the distance from the cluster centre where the 
combined cluster plus background stellar density profile is no longer readily distinguished 
from the background, as is shown in  Fig.~\ref{fig:fig1}, where we represent the mean radius 
and its error with vertical solid and dotted lines, respectively. As can be seen, 
the background level amounts in average to 25$\pm$15 per cent of the central stellar 
density, which suggests that the selected objects can be relatively easily distinguished
from the composite field population. In addition, they appear in the sky as relatively
small concentrations of stars, in average, of 1.6$\pm$0.7 arcmin in radius.

We built six CMDs and three CCDs diagrams by 
extracting every star from our $UBVRI-CT_1T_2$ photometric data sets located within 
the cluster radii estimated above. We then cleaned them from the field star
contamination by applied a statistical method which accounts for the luminosity function, 
colour distribution and stellar density of the stars distributed along the cluster 
line of sight.

The procedure was developed by \citet[see their Fig. 12]{pb12} and succefully
used elsewhere \citep[e.g.][and references therein]{p14,petal15a,petal15b,pb16}.
Briefly, it compares a extracted cluster CMD to
different field-star CMDs composed of stars located reasonably far from the
object, but not too far so as to risk losing the local field-star signature 
in terms of stellar density, luminosity function and/or colour distribution. 
The comparison between field-star and cluster CMDs is carried out by using boxes 
which vary their sizes from one place to another throughout the CMD and are 
centred on the positions of every star found in the field-star CMD. 
Note that it uses the same boxes in both cluster and field-star CMDs, and
that the comparison is performed for each individual box. 
Here we chose four field regions, each one designed to cover an
equal area as that of the cluster, and placed around the cluster. Hence,
we obtained four distint cleaned cluster CMDs. 

When comparing the four cleaned CMDs, we counted the number of times a star
remained unsubtracted in all of them. Thus, we distinguished stars that
appear once, twice, until four times, respectively. Stars appearing
once can be associated to a probability $P \le 25\%$ of being a fiducial
feature in the cleaned CMD, i.e., stars that could most frequently be found
in a field-star CMD. Stars that appear twice ($P = 50\%$) could equally 
likely be associated with either the field or the object of interest; and
those with $P \ge$ 75\%, i.e., stars found in three or four cleaned CMDs,
belong predominantly to the cleaned CMD rather than to the field-star CMDs. 
 Note that the latter can be cluster stars, or a group of stars that are
distinguishable in terms of luminosity function, 
colour distribution and stellar density without forming a physical system.
Nevertheless, the cleaning of the cluster CMDs is a mandatory step in
order to assess the real status of a catalogued aggregate and, if
confirmed as a star cluster, to estimate its astropysical properties.
Statistically speaking, a certain amount of residuals is expected, 
which depends on the degree of variability of the stellar density,
luminosity function and colour distribution of the star fields.

Figures~\ref{fig:fig2} to \ref{fig:fig16} show the whole set of CMDs and 
CC diagrams for the cluster sample that can be exploited from the present 
extensive multi-band photometry. They include every magnitude and colour 
measurements of stars located within the respective cluster radii. We 
have also incorporated to the figures
the statistical probabilities $P$ obtained above by
distinguishing stars with different colour symbols as follows: stars that
statistically belong to the field ($P \le$ 25\%, pink), stars that might belong 
to either the field or the cluster ($P =$ 50\%, light blue), and stars that 
predominantly populate the cluster region ($P \ge$ 75\%, dark blue). 
%At first glance, many of the 
%cleaned cluster CMDs (stars with $P \ge$ 75\%) 
%surprisingly do not resemble those of star clusters. Note that supposed
%cluster stars are in general brighter than $V$ $\sim$ 14.0 mag.

\begin{table*}
\caption{$UBVRI$ and $CT_1T_2$ data of stars in the field of Alessi\,14.}
\label{tab:table2}
\tiny
\begin{tabular}{@{}lcccccccccc}\hline
Star & $x$ & $y$ & $V$ & $U-B$ & $B-V$ & $V-R$ & $V-I$ &
$T_1$ & $C-T_1$ & $T_1-T_2$ \\
   & (pixel) & (pixel) & (mag)   & (mag)  & (mag)  & 
(mag)    & (mag)    & (mag)    & (mag)   & 
(mag)   \\\hline
 -- & --& --& -- & --& --& -- & --& --& -- & --  \\
   752 & 136.390 &4032.939 &  12.932    0.010  2 &   1.280   0.035  2  &  2.244 0.018 2 &  -0.353   0.001 2   & 0.360  0.010 2  & 13.163   0.015  2 &   3.044   0.002 2  &  0.830   0.010 2\\
    753 &2568.867& 4033.493 &  12.888   0.014  2 &   0.266    0.006  1 &   1.323    0.005  1 &  -0.802  0.001 2 &  -0.490   0.010 2 &  13.566   0.020 2  &  1.125    0.003 2  &  0.423   0.012  2\\
    754 & 707.266 &4034.801 &  13.801   0.007  2   & 0.257    0.027  2   & 1.615    0.020 2&   -0.673    0.001 2 &  -0.230    0.001  2  & 14.350   0.011  2  &  1.683    0.013  2  &  0.556   0.001  2\\
 -- & --& --& -- & --& --& -- & --& --& -- & -- \\
\hline
\end{tabular}

\noindent Columns list a running number per star, its $x$ and $y$ coordinates, 
the mean $V$ 
magnitude, its rms error and the number of measurements, the colours $U-B$, $B-V$,
$V-R$, $V-I$ with their respective rms errors and number of measurements, the $T_1$
magnitude with its error and number of measurements, and the $C-T_1$ and $T_1-T_2$
colours with their respective rms errors and number of measurements.
\end{table*}

\subsection{Assessment on the physical reality}

\begin{figure*}
	\includegraphics[width=\columnwidth]{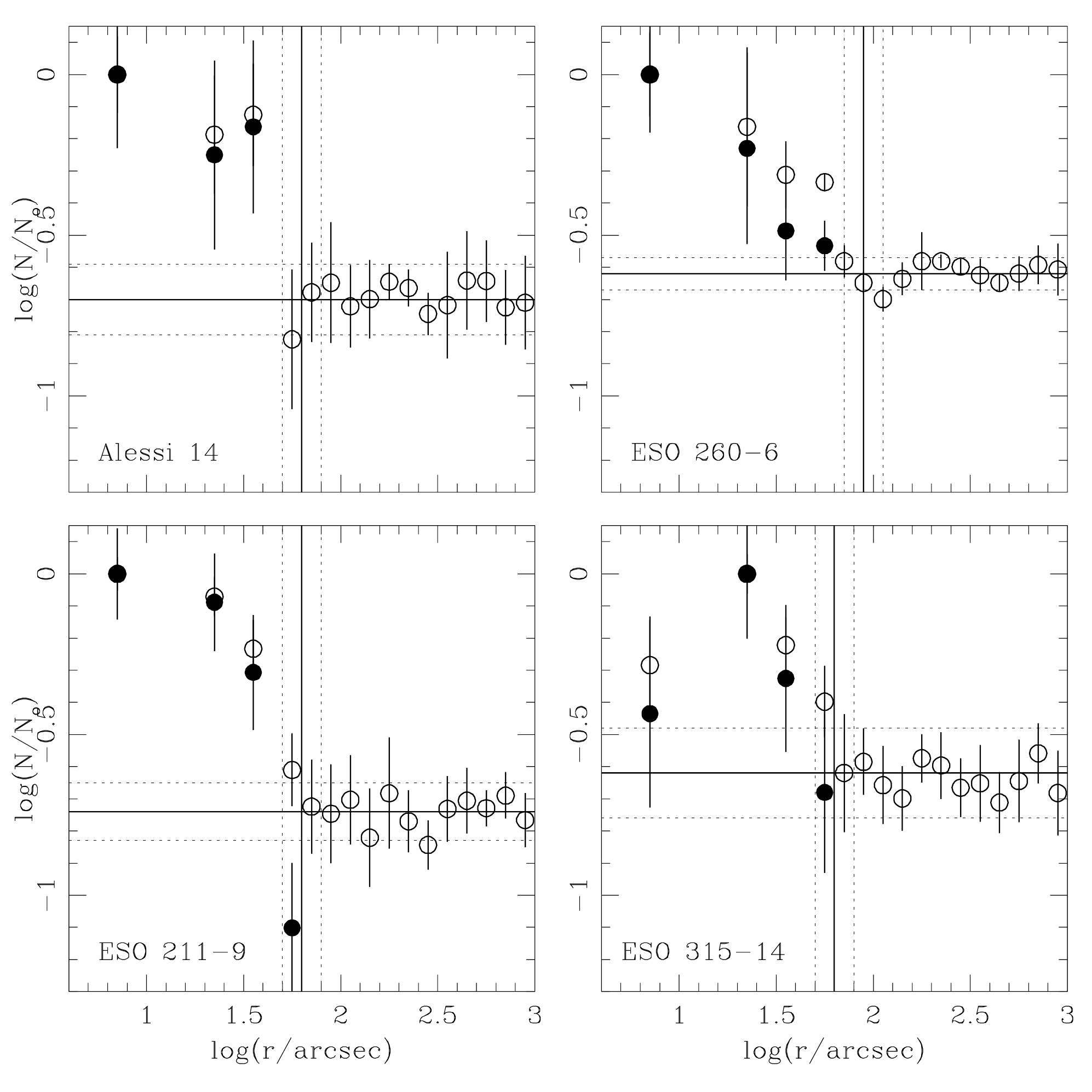}
        \includegraphics[width=\columnwidth]{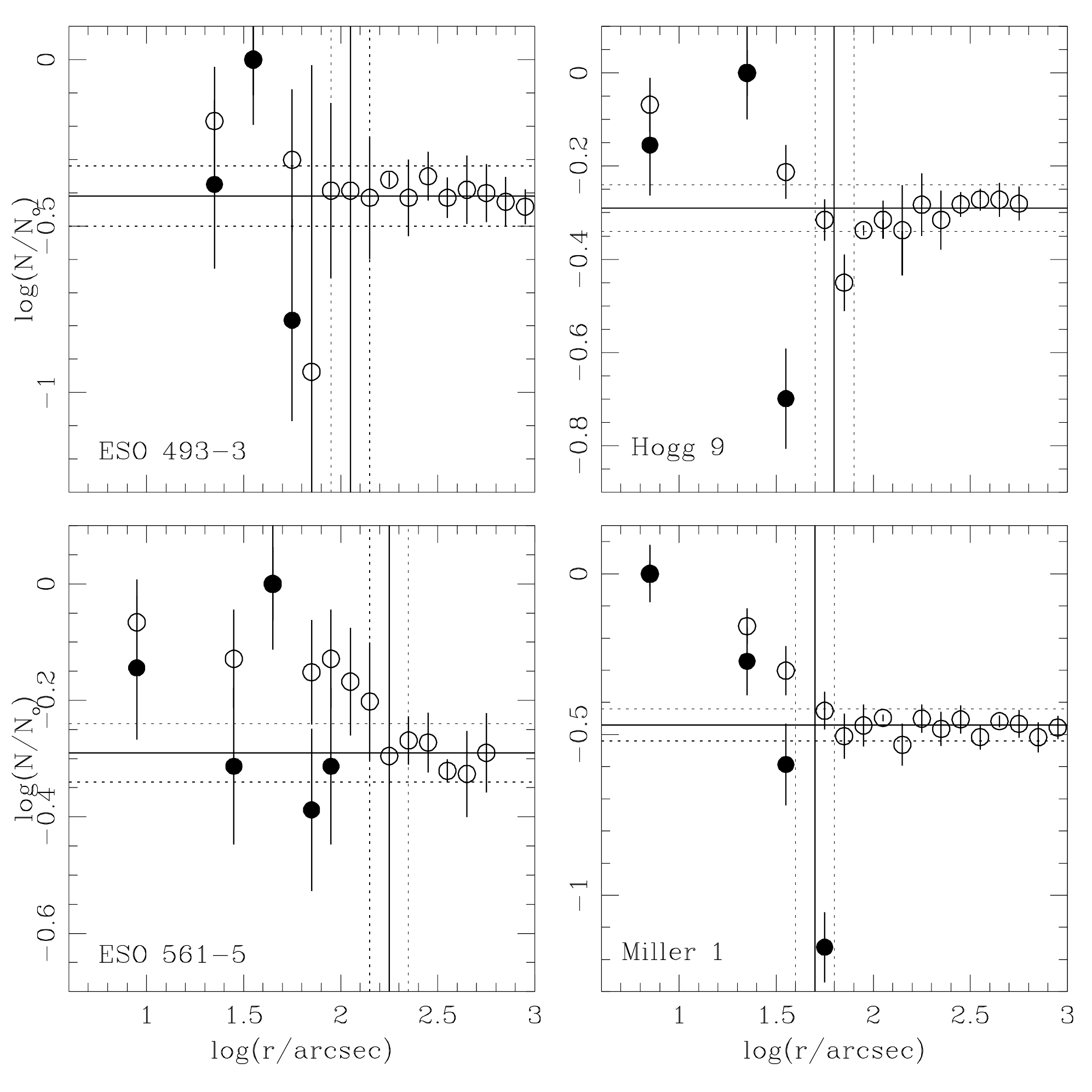}\\
	\includegraphics[width=\columnwidth]{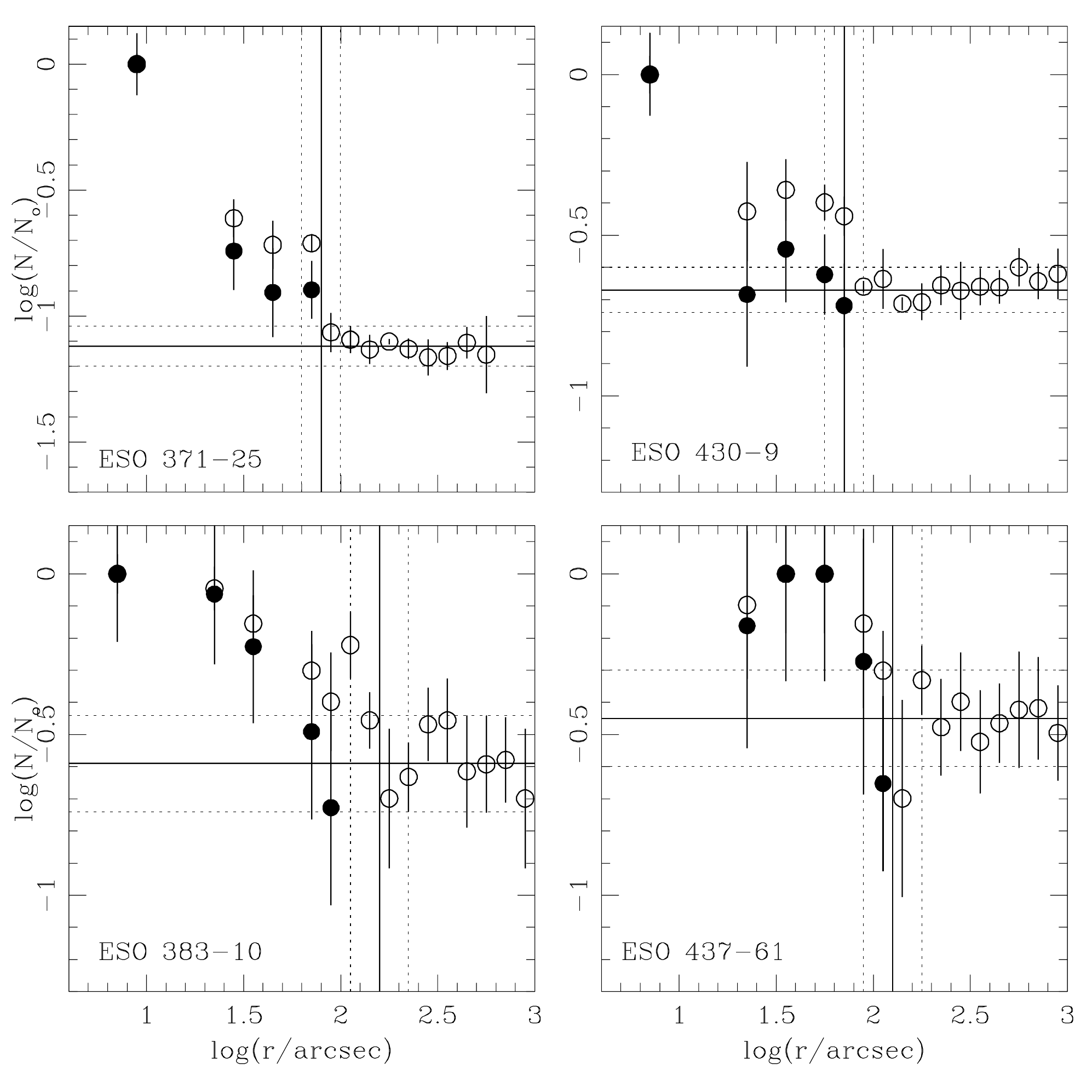}
        \includegraphics[width=\columnwidth]{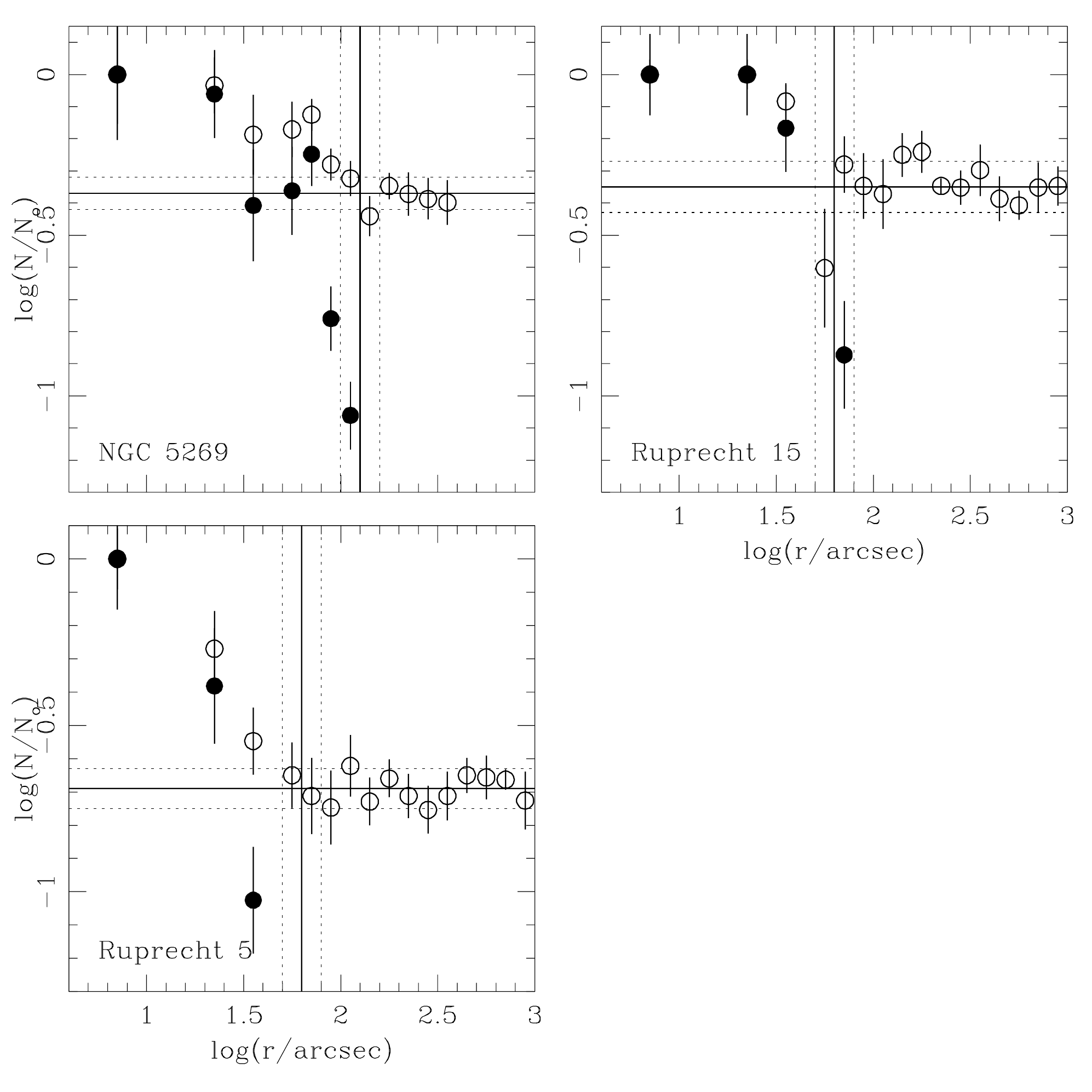}
    \caption{Stellar density profiles normalized to the central density N$_o$ obtained 
from star counts. Open and filled circles refer to measured and
background subtracted density profiles, respectively. 
%Blue and orange solid lines depict the fitted King and Plummer curves, respectively.
}
    \label{fig:fig1}
\end{figure*}

The basic challenge of trying to confirm the physical reality of the studied 
catalogued open clusters was conducted as follows: i) Identify a spatial overdensity 
relative to the background distribution (see Section 3, Fig.~\ref{fig:fig1}); ii)
 Select stars with parallaxes ($\pi$) and proper motions ($\mu$RA, $\mu$DEC) measured 
by the {\it Gaia} satellite \citep{gaia2016}. Here
we decided to dodge  the UCAC4 \citep{zachariasetal2013} catalogue, simply because its
mean proper motion errors are $\sim$ 4 mas yr$^{\rm -1}$, which are largely surpasssed
by those obtained by the {\it Gaia} mission ($\sim$ 1 mas yr$^{\rm -1}$).
Note that proper motions  distribution  observed  in an open cluster is mainly dominated 
by the observational errors \citep{diasetal2014,sa2016}.
iii) Test whether potential members, defined as those stars which comply with:

\begin{equation}
|\pi_j - \pi_k| \le |\sigma(\pi_j) + \sigma(\pi_j)|,
\end{equation}

\begin{equation}
|\mu{\rm RA}_j - \mu{\rm RA}_k| \le |\sigma(\mu{\rm RA}_j) + \sigma(\mu{\rm RA}_k)|,
\end{equation}

\begin{equation}
|\mu{\rm DEC}_j - \mu{\rm DEC}_k| \le |\sigma(\mu{\rm DEC}_j) + \sigma(\mu{\rm DEC}_k)|,
\end{equation}

\noindent where $j$ and $k$ refer to the $j$-th and $k$-th star, have 
positions in the CMDs and CC diagrams consistent with an age, distance and metallicity
within the photometric uncertainties (isochrone fitting). This last step also 
includes considering the probabilities $P$ estimated in Section 3.

We searched for parallaxes and proper motions around the central coordinates of the
studied objects and within their estimated radii using the {\it Gaia} DR1 release 
\citep{gaia2016}. The results are shown
in Table~\ref{tab:table17}, where each star is identified by a running number.
We employed the same numbering to place these stars in the
CMDs and CC diagrams (Figs.~\ref{fig:fig2} to \ref{fig:fig16}) of those objects with 
available information from {\it Gaia}. We transformed the parallaxes to true distance moduli
using the expression $(m-M)_o$ = 5$\times$$\log$(100/$\pi$).

Only ESO\,430-9 and NGC\,5269 contain stars with $P \ge$ 75\% and parallaxes in very
good agreement with the derived true distance moduli (see Section 4). As can be seen
in Figs.~\ref{fig:fig8} and \ref{fig:fig14}, both clusters
have other relatively bright stars ($V$ $\le$ 14.0 mag) -those supposed to mainly
form the stellar aggregates- with a high probability of being cluster stars ($P \ge$ 75\%)
that are pretty well matched by the same isochrone within the observable dispersion.
In the case of ESO\,315-14 (Fig.~\ref{fig:fig5}), ESO\,383-10 (Fig.~\ref{fig:fig7}) and  
ESO\,493-3 (Fig.~\ref{fig:fig10}), the bright stars' parallaxes and proper motions do not 
allow us to draw positive conclusions on the cluster reality, in good agreement with the
resulting photometric probabilities ($P \le$ 50\%, except for ESO\,315-14). Finally,
the resulting distance moduli obtained for one bright star in the field of ESO\,260-6 
(Fig.~\ref{fig:fig4}), Hogg\,9 (Fig.~\ref{fig:fig12}) and Ruprecht\,15 (Fig.~\ref{fig:fig16}),  
respectively, suggest that they are foreground stars. Additionally, since the CMDs and 
CC diagrams do not show any hint for a cluster sequence (stars with $P \ge$ 75\%),
we discard these objects as real open clusters.

For catalogued clusters without {\it Gaia} outcomes, we had to rely our analysis on the
photometric probabilities obtained in Section 3. Fortunately, ESO\,371-25 (Fig.~\ref{fig:fig6})
and ESO\,561-5 (Fig.~\ref{fig:fig11}) -with only $CT_1$ photometry- turned out to be clear 
open clusters. The latter contains only main sequence turnoff (MSTO) and red clump stars,
a signature of an advanced dynamical stage  \citep[see][]{piattietal17}; stars fainter than $T_1$ $\sim$ 16.0 mag with 
$P \ge$ 75\% (blue filled circles) are residuals from the cleaning procedure. 
ESO\,211-6  (Fig.~\ref{fig:fig3}) also shows an evident star cluster sequence. This is not
the case of Alessi\,14 (Fig.~\ref{fig:fig2}), ESO\,437-61 (Fig.~\ref{fig:fig9}), Miller\,1
(Fig.~\ref{fig:fig13}) and Ruprecht\,5 (Fig.~\ref{fig:fig15}). Their CMDs and CC diagrams
(stars with $P \ge$ 75\%) do not  simultaneously exhibit  star sequences typical of star clusters.
We recall that these objects were catalogued as open clusters because of the concentration in 
the sky of a small number of relatively bright stars, so that those stars should mainly
define the main cluster features in the CMDs and CC diagrams.

Three of the ten resulting non-physical systems, namely, ESO\,383-10, ESO\,437-61 and Hogg\,9 
have some previous relatively relevant studies. \citet{pb2007} concluded from 2MASS photometry 
and UCAC2 \citep{zachariasetal2004} proper motions that ESO\,383-10 is an open cluster remnant. 
However, the $J$ vs. $J-H$ CMD has a very narrow baseline, making field stars appear 
following a sequence similar to that of an open cluster. Such a star sequence is also seen in
Fig.~\ref{fig:fig7} -less clear in panels involving the $UB$ passbands-; it
does not resemble that of a star cluster of 2 Gyr located at 1.0 kpc from the Sun, as derived
by the authors. Moreover, the parallax of star \#2 converts to a heliocentric distance of
$\sim$ 0.3 kpc. The UCAC2 proper motion uncertainties could additionaly misled the analysis.
ESO\,437-61 resulted to be an open cluster in its late stage of dynamical evolution, according to
a star count analylis by \citet{betal01}. Later, \citet{carraroetal2005} concluded that the object
is not a physical system from the SPM3 catalogue of proper motions \citep{girardetal2004}.
Similarly, \citet{mv75} had suggested that Hogg\,9 could not be a real open cluster, although 
\citet{aetal07} assumed it as a star cluster and estimated for it an age of 300 Myr from
integrated spectroscopy.

\section{Cluster fundamental parameters}

We derived reliable ages, reddenings and distances of ESO\,211-9, ESO\,430-9 and NGC\,5269 
by making use of the six CMDs and three different CC diagrams covering wavelengths 
from the blue up to the near-infrared (Fig.~\ref{fig:fig3}, \ref{fig:fig8} and  \ref{fig:fig14},
respectively). We matched theoretical isochrones computed by \citet{betal12}
to the various CMDs and CC diagrams, simultaneosly.

As a starting point we took advantage of the shape of the main sequences (MS), its curvature 
(less and more pronounced), 
the relative distance between the giant stars and the MSTOs in magnitude and colour separately, 
among others, to find the age of the isochrone which best matches the cluster's features in the 
CMDs and CC diagrams, regardless the cluster reddening and distance.
From our best choice, we derived the cluster 
reddenings by shifting that isochrone in the three CC diagrams
following the reddening vectors until their bluest points coincided with the observed ones.
Note that this requirement allowed us to use the $V-R$ vs $R-I$ CC diagram as well, even 
though the reddening vector runs almost parallell to the cluster sequence.
In order to enter the isochrones into the CMDs and CC diagrams we 
used the following ratios: $E(U-B)$/$E(B-V)$ = 0.72 + 0.05$\times$$E(B-V)$ \citep{hj56}; 
$E(V-R)$/$E(B-V)$ = 0.65, $E(V-I)$/$E(B-V)$ = 1.25, $A_{V}$/$E(B-V)$ = 3.1 \citep{cetal89}; 
$E(C-T_1)$/$E(B-V)$ = 1.97, $E(T_1-T_2)$/$E(B-V)$ = 0.692, $A_{T_1}$/$E(B-V)$ = 2.62 
\citep{g96}. Finally, the adopted $E(B-V)$ colour excesses were used to properly shift the 
chosen isochrone in the CMDs in order to derive the cluster true distance moduli 
$(m-M)_o$ by shifting the isochrone along the magnitude axes. We used isochrones
of solar metal content ([Fe/H] = 0.0 dex) and checked that those for [Fe/H] = $\pm$0.2 dex
did not produce any visible difference in the CMDs and CC diagrams. Lower metallicities
values  (-0.7 $\le$ [Fe/H] $\le$ -0.2, \citep{paunzeretal2010,hetal14}) did not match
the CMDs and CC diagrams satisfactority.

\setcounter{table}{16}
\begin{table}
\caption{{\it Gaia} DR1 parallaxes and proper motions for stars around the studied
cluster fields.}
\label{tab:table17}
\begin{tabular}{@{}lcccc}\hline
   &  ID  & $\pi$ & $\mu${\rm RA} & $\mu${\rm DEC} \\
   &      & (mas) &    (mas yr$^{\rm -1}$) &  (mas yr$^{\rm -1}$)\\\hline
ESO\,260-6    & 1 &  1.85$\pm$0.51 & 7.06$\pm$2.23  &-19.89$\pm$1.30\\
              &   &                &                &              \\
ESO\,315-14   & 1 & 2.27$\pm$0.68  &-12.05$\pm$1.33 &-1.08$\pm$2.19\\
              & 2 & 0.83$\pm$0.33  &-6.59$\pm$0.99  & 3.19$\pm$1.07\\
              & 3 & 2.06$\pm$0.34  &-8.28$\pm$0.99  &-3.59$\pm$1.03\\
              &   &                &                &              \\
ESO\,383-10   & 1 & 2.33$\pm$0.54  &-19.97$\pm$2.01 &-25.08$\pm$0.68\\
              & 2 & 3.46$\pm$0.54  & 3.79$\pm$1.50  &-6.63$\pm$0.44\\
              &   &                &                &              \\
ESO\,430-9    & 1 & 0.51$\pm$0.26  &-2.45$\pm$0.71  & 2.17$\pm$0.73\\
              &   &                &                &              \\
ESO\,493-3    & 1 & 0.78$\pm$0.32  &-3.45$\pm$0.97  & 1.09$\pm$1.17\\
              & 2 &-0.14$\pm$0.92  &-2.90$\pm$2.55  & 1.56$\pm$2.81\\
              & 3 & 2.54$\pm$0.26  &-1.48$\pm$0.91  &32.27$\pm$1.05\\
              & 4 & 0.62$\pm$0.30  &-3.86$\pm$0.91  & 3.00$\pm$1.13\\
              & 5 & 1.95$\pm$0.29  &-14.96$\pm$0.91 &-0.66$\pm$1.07\\
              & 6 & 1.62$\pm$0.26  & 3.40$\pm$0.78  &-9.54$\pm$0.96\\
              & 7 & 2.89$\pm$0.33  &-10.81$\pm$0.97 & 20.59$\pm$1.02\\
              &   &                &                &              \\
Hogg\,9       & 1 & 1.23$\pm$0.44  &-4.50$\pm$1.98  & 1.83$\pm$0.93\\
              &   &                &                &              \\
NGC\,5269     & 1 & 0.37$\pm$0.29  &-4.58$\pm$0.80  &-1.79$\pm$0.68\\
              & 2 & 0.31$\pm$0.27  &-4.75$\pm$0.72  &-2.04$\pm$0.61\\
              & 3 & 0.51$\pm$0.24  &-4.44$\pm$0.63  &-2.44$\pm$0.55\\
              & 4 & 0.39$\pm$0.25  &-4.83$\pm$0.66  &-2.26$\pm$0.57\\
              &   &                &                &              \\
Ruprecht\,15  & 1 & 0.25$\pm$0.47 &-2.02$\pm$1.07   & 0.18$\pm$1.02\\

\hline
\end{tabular}
\end{table}

For ESO\,371-25 and ESO\,561-5 -with only $CT_1$ photometry- we made use of the age-metallicity 
diagnostic diagram, $\delta$$T_1$ versus
$\delta$$C$ - $\delta$$T_1$, which has shown the ability of unambiguously providing 
age and metallicity estimates,  simultaneously \citep[][see their Fig. 4]{pp15}. 
$\delta$$C$ and $\delta$$T_1$ are 
the respective magnitude differences between the giant red clump (RC) and the MSTO.
The diagnostic diagram allows to derive ages from 1 up to 13 Gyr and metallicities [Fe/H] 
from -2.0 up to +0.5 dex, and is independent of the cluster reddening and distance modulus. 
We measured $C$ and $T_1$ magnitudes at the MSTO and RC, then computed 
$\delta$$C$ and $\delta$$T_1$ and entered into Figure 4 of \citet{pp15}  to estimate 
cluster ages and metallicities (both clusters are of solar metal content within
$\sigma$[Fe/H] = 0.15 dex). Then, we derived the cluster reddening and distance moduli 
by matching the respective isochrones to the cluster CMDs.
%We found that isochrones bracketing the age choiced  by $\Delta$
%log($t$ yr$^{-1}$) = $\pm$0.10 and $\Delta$[Fe/H] = $\pm$0.10 dex represent the overall 
%age/metallicity uncertainties
%owing to the observed dispersion in the cluster CMDs and CC diagrams. 
The adopted best matched isochrones for ESO\,211-9, ESO\,371-25, ESO\,430-9, ESO\,561-5 
and NGC\,5269 are overplotted with black solid lines on Fig.~\ref{fig:fig3}, \ref{fig:fig6} 
\ref{fig:fig8}, \ref{fig:fig11} and  \ref{fig:fig14}, respectively, 
while the resulting values with their errors for the cluster reddenings, distances, 
ages are listed in Table~\ref{tab:table18}.

Fig.~\ref{fig:fig17} depicts the spatial distribution of the studied clusters, where
we added for comparison purposes the 2167 open clusters 
catalogued by 
\citet[][version 3.4 as of January 2016]{detal02} and the schematic positions of the spiral
arms \citep{ds2001,moitinhoetal2006}. The cluster Galactic coordinates were computed
 using the derived cluster heliocentric distances, their angular
Galactic coordinates and a Galactocentric distance of the Sun of R$_{GC_\odot}$ = 8.3 kpc 
\citep[][and references therein]{hh2014}. The clusters are
distributed outside the circle around the Sun 
(d $\sim$ 2.0 kpc) where the catalogued clusters are mostly concentrated. 
They belong to the thin Galactic disc, with the exception of ESO\,371-25 
which is nearly 0.5 kpc above the Galactic plane. The relationship between
their positions in the Galaxy and their ages are in agreement with the overall
feature of the Galactic disc attained from open cluster statistical analyses
\citep[see, e.g.][]{bonattoetal2006,joshietal2016}.

\begin{table}
\centering
\caption{Derived properties of selected open clusters.}
\label{tab:table18}
\begin{tabular}{@{}lccccccccccc}\hline
Star cluster & $E(B-V)$ & $(m-M)_o$ & d  & $\log(t)$ \\
             &  (mag)   & (mag)     & (kpc) &           \\\hline
ESO\,211-9   &  1.40$\pm$0.10 & 12.0$\pm$0.3 & 2.50$\pm$0.35  & 8.70$\pm$0.10  \\
ESO\,371-25  &  0.40$\pm$0.05 & 13.7$\pm$0.2 & 5.49$\pm$1.25  & 9.40$\pm$0.05  \\
ESO\,430-9   &  0.45$\pm$0.05 & 12.2$\pm$0.3 & 2.75$\pm$0.38  & 8.00$\pm$0.10   \\
ESO\,561-5   &  0.25$\pm$0.05 & 11.6$\pm$0.2 & 2.09$\pm$0.19  & 8.90$\pm$0.05   \\
NGC\,5269    &  0.25$\pm$0.05 & 11.5$\pm$0.3 & 1.99$\pm$0.27  & 8.50$\pm$0.10  \\   
\hline
\end{tabular}
%\noindent Note: to convert 1 arcsec to pc, we use the following expression,10$\times$10$^{(m-M)_o/5}$sin(1/3600)
%where $(m-M)_o$ is the true distance modulus.

\end{table}

\setcounter{figure}{16}
\begin{figure}
	\includegraphics[width=\columnwidth]{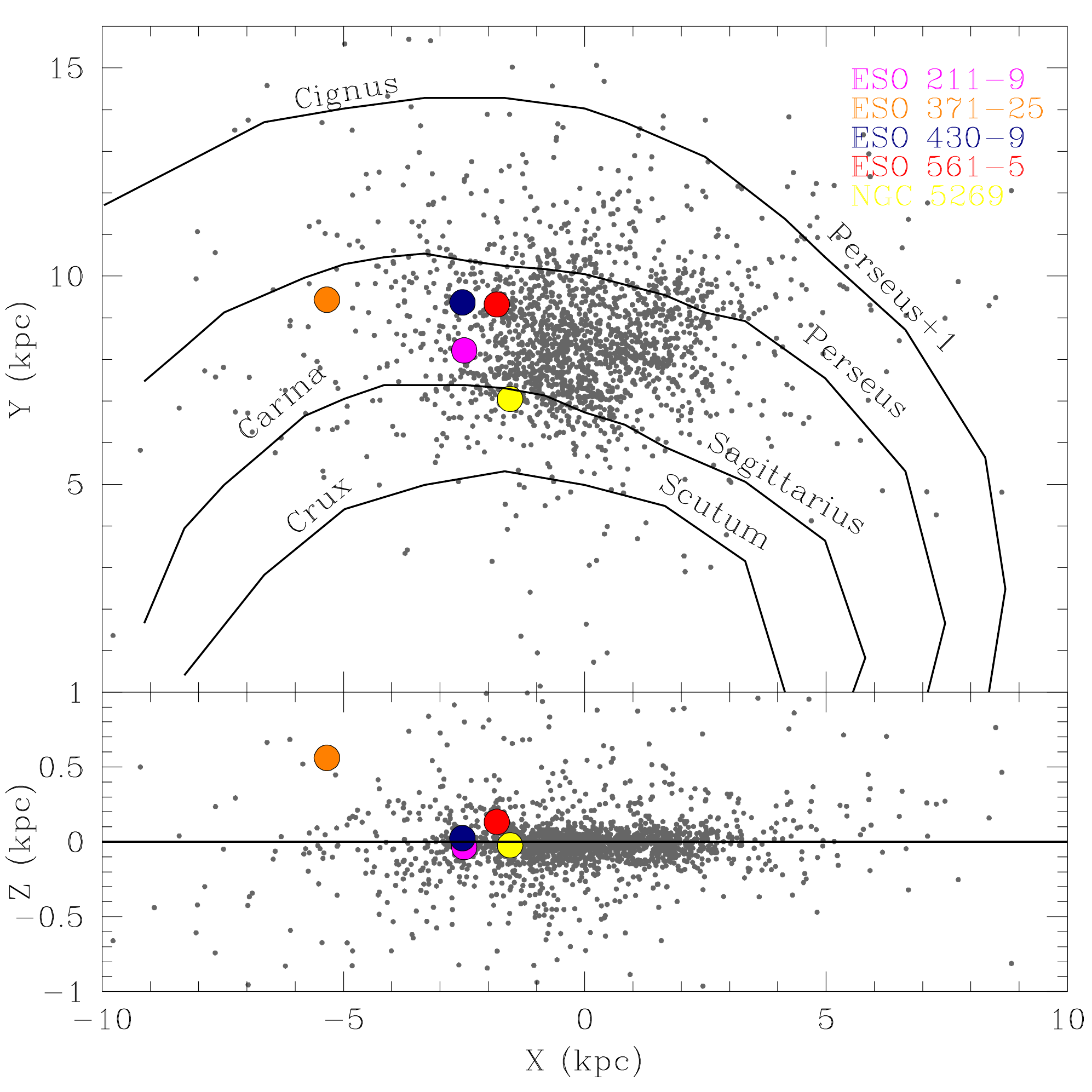}
    \caption{Galactic spatial distribution of the studied clusters. Open clusters from the
catalogue of \citet[][version 3.4 as of January 2016]{detal02} are drawn with gray dots, while
the schematic positions of spiral arms \citep{ds2001,moitinhoetal2006} are traced with black 
solid lines.}
    \label{fig:fig17}
\end{figure}

\section{conclusions}

With the aim of improving the statistics of well studied open clusters, 
we present a multi-band photometric analysis of 15 catalogued open clusters.
The selected sample was chosen on the basis of the relative high brightness and
compact appearance of the candidates and because they had not been studied or poorly 
studied at the time of designing the observational campaign.

We used publicly available Johnson $UBV$,  Kron-Cousins $RI$ and Washington $C$ 
images obtained at CTIO to produce photometric data sets from which we built
 six CMDs and three different CC diagrams per object covering wavelengths 
from the blue up to the near-infrared. This unprecedented multi-band coverage
allowed us to thoroughly judged their reality as stellar aggregates.

In order to reliably assess on their physical nature we statistically assigned
to each observed star in a given field a probability of being a fiducial feature
of that field in terms of its local luminosity function, colour distribution and 
stellar density. This was accomplished by using a powerful technique that makes 
use of cells varying in position and size in the CMDs to disentangle the
fiducial from the contaminating field stars. To that purpuse we first traced the
cluster stellar radial profiles and estimated their radii that were used to
delimit the region to examine. Likewise, we took advantage of accurate parallaxes
and proper motions measured by the {\it Gaia} satellite to help our
decision on the open cluster reality. 

Contrarily to what could be expected, as judged from the objects' brightness and 
compactness, 10 catalogued aggregates did not show any hint of being real physical systems.  
Three of them (ESO\,383-10, ESO\,437-61 and Hogg\,9) had been treated as open 
clusters in previous photometric studies. Thus, this result points to the need of
a thorough examination of any group of stars believed to be a physical entity by
its only appearance in the sky. On the other hand, we estimated reliable fundamental
parameters for the remaining 5 studied objects, which were confirmed as real
open clusters. By exploiting the wealth of photometric data sets in combination with theoretical
isochrones, we found out that the clusters in our sample are in wide age range, 
8.0 $\le$ $\log$($t$ yr$^{-1}$) $\le$ 9.4, of solar metal content and placed between 2.0 and 5.5 kpc
from the Sun. They resulting astrophysical properties are in agreement with the updated
known picture of the spatial distribution of ages/metallicities of open clusters in the
Galactic disc.

\section*{Acknowledgements}
This work has made use of data from the European Space Agency (ESA)
mission {\it Gaia} (\url{http://www.cosmos.esa.int/gaia}), processed by
the {\it Gaia} Data Processing and Analysis Consortium (DPAC,
\url{http://www.cosmos.esa.int/web/gaia/dpac/consortium}). Funding
for the DPAC has been provided by national institutions, in particular
the institutions participating in the {\it Gaia} Multilateral Agreement.
 We thank the anonymous referee whose thorough comments and suggestions
allowed us to improve the manuscript.

%%%%%%%%%%%%%%%%%%%%%%%%%%%%%%%%%%%%%%%%%%%%%%%%%%



\bibliographystyle{mnras}
%\bibliography{paper} % if your bibtex file is called paper.bib

\begin{thebibliography}{}
\makeatletter
\relax
\def\mn@urlcharsother{\let\do\@makeother \do\$\do\&\do\#\do\^\do\_\do\%\do\~}
\def\mn@doi{\begingroup\mn@urlcharsother \@ifnextchar [ {\mn@doi@}
  {\mn@doi@[]}}
\def\mn@doi@[#1]#2{\def\@tempa{#1}\ifx\@tempa\@empty \href
  {http://dx.doi.org/#2} {doi:#2}\else \href {http://dx.doi.org/#2} {#1}\fi
  \endgroup}
\def\mn@eprint#1#2{\mn@eprint@#1:#2::\@nil}
\def\mn@eprint@arXiv#1{\href {http://arxiv.org/abs/#1} {{\tt arXiv:#1}}}
\def\mn@eprint@dblp#1{\href {http://dblp.uni-trier.de/rec/bibtex/#1.xml}
  {dblp:#1}}
\def\mn@eprint@#1:#2:#3:#4\@nil{\def\@tempa {#1}\def\@tempb {#2}\def\@tempc
  {#3}\ifx \@tempc \@empty \let \@tempc \@tempb \let \@tempb \@tempa \fi \ifx
  \@tempb \@empty \def\@tempb {arXiv}\fi \@ifundefined
  {mn@eprint@\@tempb}{\@tempb:\@tempc}{\expandafter \expandafter \csname
  mn@eprint@\@tempb\endcsname \expandafter{\@tempc}}}

\bibitem[\protect\citeauthoryear{{Ahumada}, {Clari{\'a}}  \& {Bica}}{{Ahumada}
  et~al.}{2007}]{aetal07}
{Ahumada} A.~V.,  {Clari{\'a}} J.~J.,   {Bica} E.,  2007, \mn@doi [\aap]
  {10.1051/0004-6361:20077608}, 473, 437

\bibitem[\protect\citeauthoryear{{Bica}, {Santiago}, {Dutra}, {Dottori}, {de
  Oliveira}  \& {Pavani}}{{Bica} et~al.}{2001}]{betal01}
{Bica} E.,  {Santiago} B.~X.,  {Dutra} C.~M.,  {Dottori} H.,  {de Oliveira}
  M.~R.,   {Pavani} D.,  2001, \mn@doi [\aap] {10.1051/0004-6361:20000248},
  366, 827

\bibitem[\protect\citeauthoryear{{Bonatto}, {Kerber}, {Bica}  \&
  {Santiago}}{{Bonatto} et~al.}{2006}]{bonattoetal2006}
{Bonatto} C.,  {Kerber} L.~O.,  {Bica} E.,   {Santiago} B.~X.,  2006, \mn@doi
  [\aap] {10.1051/0004-6361:20053573}, \href
  {http://adsabs.harvard.edu/abs/2006A%26A...446..121B} {446, 121}

\bibitem[\protect\citeauthoryear{{Bressan}, {Marigo}, {Girardi}, {Salasnich},
  {Dal Cero}, {Rubele}  \& {Nanni}}{{Bressan} et~al.}{2012}]{betal12}
{Bressan} A.,  {Marigo} P.,  {Girardi} L.,  {Salasnich} B.,  {Dal Cero} C.,
  {Rubele} S.,   {Nanni} A.,  2012, \mn@doi [\mnras]
  {10.1111/j.1365-2966.2012.21948.x}, 427, 127

\bibitem[\protect\citeauthoryear{{Cardelli}, {Clayton}  \& {Mathis}}{{Cardelli}
  et~al.}{1989}]{cetal89}
{Cardelli} J.~A.,  {Clayton} G.~C.,   {Mathis} J.~S.,  1989, \mn@doi [\apj]
  {10.1086/167900}, 345, 245

\bibitem[\protect\citeauthoryear{{Carraro}, {Dinescu}, {Girard}  \& {van
  Altena}}{{Carraro} et~al.}{2005}]{carraroetal2005}
{Carraro} G.,  {Dinescu} D.~I.,  {Girard} T.~M.,   {van Altena} W.~F.,  2005,
  \mn@doi [\aap] {10.1051/0004-6361:20042044}, \href
  {http://adsabs.harvard.edu/abs/2005A%26A...433..143C} {433, 143}

\bibitem[\protect\citeauthoryear{{Collinder}}{{Collinder}}{1931}]{collinder1931}
{Collinder} P.,  1931, Annals of the Observatory of Lund, \href
  {http://adsabs.harvard.edu/abs/1931AnLun...2....1C} {2, B1}

\bibitem[\protect\citeauthoryear{{Dias}, {Alessi}, {Moitinho}  \&
  {L{\'e}pine}}{{Dias} et~al.}{2002}]{detal02}
{Dias} W.~S.,  {Alessi} B.~S.,  {Moitinho} A.,   {L{\'e}pine} J.~R.~D.,  2002,
  \mn@doi [\aap] {10.1051/0004-6361:20020668}, 389, 871

\bibitem[\protect\citeauthoryear{{Dias}, {Monteiro}, {Caetano}, {L{\'e}pine},
  {Assafin}  \& {Oliveira}}{{Dias} et~al.}{2014}]{diasetal2014}
{Dias} W.~S.,  {Monteiro} H.,  {Caetano} T.~C.,  {L{\'e}pine} J.~R.~D.,
  {Assafin} M.,   {Oliveira} A.~F.,  2014, \mn@doi [\aap]
  {10.1051/0004-6361/201323226}, \href
  {http://adsabs.harvard.edu/abs/2014A%26A...564A..79D} {564, A79}

\bibitem[\protect\citeauthoryear{{Drimmel} \& {Spergel}}{{Drimmel} \&
  {Spergel}}{2001}]{ds2001}
{Drimmel} R.,  {Spergel} D.~N.,  2001, \mn@doi [\apj] {10.1086/321556}, \href
  {http://adsabs.harvard.edu/abs/2001ApJ...556..181D} {556, 181}

\bibitem[\protect\citeauthoryear{{Gaia Collaboration}, {Brown}, {Vallenari},
  {Prusti}, {de Bruijne}, {Mignard}, {Drimmel}  \& {co-authors}}{{Gaia
  Collaboration} et~al.}{2016}]{gaia2016}
{Gaia Collaboration} {Brown} A.~G.~A.,  {Vallenari} A.,  {Prusti} T.,  {de
  Bruijne} J.,  {Mignard} F.,  {Drimmel} R.,   {co-authors} .,  2016, preprint,
  \href {http://adsabs.harvard.edu/abs/2016arXiv160904172G} {} (\mn@eprint
  {arXiv} {1609.04172})

\bibitem[\protect\citeauthoryear{{Geisler}}{{Geisler}}{1996}]{g96}
{Geisler} D.,  1996, \mn@doi [\aj] {10.1086/117799}, 111, 480

\bibitem[\protect\citeauthoryear{{Girard}, {Dinescu}, {van Altena}, {Platais},
  {Lopez}  \& {Monet}}{{Girard} et~al.}{2004}]{girardetal2004}
{Girard} T.~M.,  {Dinescu} D.~I.,  {van Altena} W.~F.,  {Platais} I.,  {Lopez}
  C.~E.,   {Monet} D.~G.,  2004, in {Clemens} D.,  {Shah} R.,   {Brainerd} T.,
  eds,  Astronomical Society of the Pacific Conference Series Vol. 317, Milky
  Way Surveys: The Structure and Evolution of our Galaxy. p.~206

\bibitem[\protect\citeauthoryear{{Heiter}, {Soubiran}, {Netopil}  \&
  {Paunzen}}{{Heiter} et~al.}{2014}]{hetal14}
{Heiter} U.,  {Soubiran} C.,  {Netopil} M.,   {Paunzen} E.,  2014, \mn@doi
  [\aap] {10.1051/0004-6361/201322559}, 561, A93

\bibitem[\protect\citeauthoryear{{Hiltner} \& {Johnson}}{{Hiltner} \&
  {Johnson}}{1956}]{hj56}
{Hiltner} W.~A.,  {Johnson} H.~L.,  1956, \mn@doi [\apj] {10.1086/146231},
  \href {http://adsabs.harvard.edu/abs/1956ApJ...124..367H} {124, 367}

\bibitem[\protect\citeauthoryear{{Hou} \& {Han}}{{Hou} \& {Han}}{2014}]{hh2014}
{Hou} L.~G.,  {Han} J.~L.,  2014, \mn@doi [\aap] {10.1051/0004-6361/201424039},
  \href {http://adsabs.harvard.edu/abs/2014A%26A...569A.125H} {569, A125}

\bibitem[\protect\citeauthoryear{{Joshi}, {Dambis}, {Pandey}  \&
  {Joshi}}{{Joshi} et~al.}{2016}]{joshietal2016}
{Joshi} Y.~C.,  {Dambis} A.,  {Pandey} A.~K.,   {Joshi} S.,  2016, preprint,
  \href {http://adsabs.harvard.edu/abs/2016arXiv160606425J} {} (\mn@eprint
  {arXiv} {1606.06425})

\bibitem[\protect\citeauthoryear{{Moffat} \& {Vogt}}{{Moffat} \&
  {Vogt}}{1975}]{mv75}
{Moffat} A.~F.~J.,  {Vogt} N.,  1975, \aaps, \href
  {http://adsabs.harvard.edu/abs/1975A%26AS...20..125M} {20, 125}

\bibitem[\protect\citeauthoryear{{Moitinho}, {V{\'a}zquez}, {Carraro}, {Baume},
  {Giorgi}  \& {Lyra}}{{Moitinho} et~al.}{2006}]{moitinhoetal2006}
{Moitinho} A.,  {V{\'a}zquez} R.~A.,  {Carraro} G.,  {Baume} G.,  {Giorgi}
  E.~E.,   {Lyra} W.,  2006, \mn@doi [\mnras]
  {10.1111/j.1745-3933.2006.00163.x}, \href
  {http://adsabs.harvard.edu/abs/2006MNRAS.368L..77M} {368, L77}

\bibitem[\protect\citeauthoryear{{Paunzen}, {Heiter}, {Netopil}  \&
  {Soubiran}}{{Paunzen} et~al.}{2010}]{paunzeretal2010}
{Paunzen} E.,  {Heiter} U.,  {Netopil} M.,   {Soubiran} C.,  2010, \mn@doi
  [\aap] {10.1051/0004-6361/201014131}, \href
  {http://adsabs.harvard.edu/abs/2010A%26A...517A..32P} {517, A32}

\bibitem[\protect\citeauthoryear{{Pavani} \& {Bica}}{{Pavani} \&
  {Bica}}{2007}]{pb2007}
{Pavani} D.~B.,  {Bica} E.,  2007, \mn@doi [\aap] {10.1051/0004-6361:20066240},
  \href {http://adsabs.harvard.edu/abs/2007A%26A...468..139P} {468, 139}

\bibitem[\protect\citeauthoryear{{Piatti}}{{Piatti}}{2014}]{p14}
{Piatti} A.~E.,  2014, \mn@doi [\mnras] {10.1093/mnras/stu534}, 440, 3091

\bibitem[\protect\citeauthoryear{{Piatti}}{{Piatti}}{2016}]{piatti16b}
{Piatti} A.~E.,  2016, \mn@doi [\mnras] {10.1093/mnras/stw2248}, \href
  {http://adsabs.harvard.edu/abs/2016MNRAS.463.3476P} {463, 3476}

\bibitem[\protect\citeauthoryear{{Piatti} \& {Bastian}}{{Piatti} \&
  {Bastian}}{2016}]{pb16}
{Piatti} A.~E.,  {Bastian} N.,  2016, \mn@doi [\aap]
  {10.1051/0004-6361/201628339}, \href
  {http://adsabs.harvard.edu/abs/2016A%26A...590A..50P} {590, A50}

\bibitem[\protect\citeauthoryear{{Piatti} \& {Bica}}{{Piatti} \&
  {Bica}}{2012}]{pb12}
{Piatti} A.~E.,  {Bica} E.,  2012, \mn@doi [\mnras]
  {10.1111/j.1365-2966.2012.21694.x}, 425, 3085

\bibitem[\protect\citeauthoryear{{Piatti} \& {Perren}}{{Piatti} \&
  {Perren}}{2015}]{pp15}
{Piatti} A.~E.,  {Perren} G.~I.,  2015, \mn@doi [\mnras]
  {10.1093/mnras/stv861}, \href
  {http://adsabs.harvard.edu/abs/2015MNRAS.450.3771P} {450, 3771}

\bibitem[\protect\citeauthoryear{{Piatti}, {Clari{\'a}}, {Parisi}  \&
  {Ahumada}}{{Piatti} et~al.}{2011}]{piattietal2011}
{Piatti} A.~E.,  {Clari{\'a}} J.~J.,  {Parisi} M.~C.,   {Ahumada} A.~V.,  2011,
  \mn@doi [\pasp] {10.1086/659848}, \href
  {http://adsabs.harvard.edu/abs/2011PASP..123..519P} {123, 519}

\bibitem[\protect\citeauthoryear{{Piatti}, {de Grijs}, {Rubele}, {Cioni},
  {Ripepi}  \& {Kerber}}{{Piatti} et~al.}{2015a}]{petal15a}
{Piatti} A.~E.,  {de Grijs} R.,  {Rubele} S.,  {Cioni} M.-R.~L.,  {Ripepi} V.,
   {Kerber} L.,  2015a, \mn@doi [\mnras] {10.1093/mnras/stv635}, 450, 552

\bibitem[\protect\citeauthoryear{{Piatti} et~al.,}{{Piatti}
  et~al.}{2015b}]{petal15b}
{Piatti} A.~E.,  et~al., 2015b, \mn@doi [\mnras] {10.1093/mnras/stv2054}, \href
  {http://adsabs.harvard.edu/abs/2015MNRAS.454..839P} {454, 839}

\bibitem[\protect\citeauthoryear{{Piatti}, {Dias}  \& {Sampedro}}{{Piatti}
  et~al.}{2017}]{piattietal17}
{Piatti} A.~E.,  {Dias} W.~S.,   {Sampedro} L.~M.,  2017, \mnrassub

\bibitem[\protect\citeauthoryear{{Ruprecht}}{{Ruprecht}}{1966}]{ruprecht1966}
{Ruprecht} J.,  1966, Bulletin of the Astronomical Institutes of
  Czechoslovakia, \href {http://adsabs.harvard.edu/abs/1966BAICz..17...33R}
  {17, 33}

\bibitem[\protect\citeauthoryear{{Sampedro} \& {Alfaro}}{{Sampedro} \&
  {Alfaro}}{2016}]{sa2016}
{Sampedro} L.,  {Alfaro} E.~J.,  2016, \mn@doi [\mnras] {10.1093/mnras/stw243},
  \href {http://adsabs.harvard.edu/abs/2016MNRAS.457.3949S} {457, 3949}

\bibitem[\protect\citeauthoryear{{Zacharias}, {Urban}, {Zacharias}, {Wycoff},
  {Hall}, {Monet}  \& {Rafferty}}{{Zacharias} et~al.}{2004}]{zachariasetal2004}
{Zacharias} N.,  {Urban} S.~E.,  {Zacharias} M.~I.,  {Wycoff} G.~L.,  {Hall}
  D.~M.,  {Monet} D.~G.,   {Rafferty} T.~J.,  2004, \mn@doi [\aj]
  {10.1086/386353}, \href {http://adsabs.harvard.edu/abs/2004AJ....127.3043Z}
  {127, 3043}

\bibitem[\protect\citeauthoryear{{Zacharias}, {Finch}, {Girard}, {Henden},
  {Bartlett}, {Monet}  \& {Zacharias}}{{Zacharias}
  et~al.}{2013}]{zachariasetal2013}
{Zacharias} N.,  {Finch} C.~T.,  {Girard} T.~M.,  {Henden} A.,  {Bartlett}
  J.~L.,  {Monet} D.~G.,   {Zacharias} M.~I.,  2013, \mn@doi [\aj]
  {10.1088/0004-6256/145/2/44}, \href
  {http://adsabs.harvard.edu/abs/2013AJ....145...44Z} {145, 44}

\makeatother
\end{thebibliography}

%to be uncommented before sending to editor
\input{paper.bbl}

%\begin{thebibliography}{99}
%\end{thebibliography}

\setcounter{figure}{1}
\begin{landscape}
\begin{figure} 
\caption{CMDs and CC diagrams for stars measured in the field of Alessi\,14.
Colour-scaled symbols represent stars with photometric memberships 
$P \le$ 25\% (pink), equals to 50\% (light blue) and $\ge$ 75\% (dark
blue), respectively. 
%We overplotted the isochrones which best matches the
%cluster features (black solid line) and that for the CE solution (red line),
%respectively. 
}
   \label{fig:fig2}
\end{figure}
\end{landscape}

\begin{landscape}
\begin{figure} 
\caption{CMDs and CC diagrams for stars measured in the field of ESO\,211-6. Symbols
are as in Fig.~\ref{fig:fig2}. We overplotted the isochrone which best matches the
cluster features (black solid line). In panels involving $U-B$ colours, the isochrone
was drawn for complete purposes.}
   \label{fig:fig3}
\end{figure}
\end{landscape}

\begin{landscape}
\begin{figure} 
\caption{CMDs and CC diagrams for stars measured in the field of ESO\,260-6. Symbols
are as in Fig.~\ref{fig:fig2}. Star with {\it Gaia} parallax and
proper motions is numbered \#1 . (see text for details).}
   \label{fig:fig4}
\end{figure}
\end{landscape}

\begin{landscape}
\begin{figure} 
\caption{CMDs and CC diagrams for stars measured in the field of ESO\,315-14. Symbols
are as in Fig.~\ref{fig:fig2}.  Stars with {\it Gaia} parallaxes and
proper motions are numbered from \#1 to 3. (see text for details).}
\label{fig:fig5}
\end{figure}
\end{landscape}

\begin{figure} 
\caption{CMD for stars measured in the field of ESO\,371-25. Symbols
are as in Fig.~\ref{fig:fig2}.  We overplotted the isochrone which best matches the
cluster features (black solid line). }
\label{fig:fig6}
\end{figure}

\begin{landscape}
\begin{figure} 
\caption{CMDs and CC diagrams for stars measured in the field of ESO\,383-10. Symbols
are as in Fig.~\ref{fig:fig2}. Stars with {\it Gaia} parallaxes and
proper motions are numbered from \#1 to 2. (see text for details).}
\label{fig:fig7}
\end{figure}
\end{landscape}

\begin{landscape}
\begin{figure} 
\caption{CMDs and CC diagrams for stars measured in the field of ESO\,430-9. Symbols
are as in Fig.~\ref{fig:fig2}. Star with {\it Gaia} parallax and
proper motions is numbered \#1 (see text for details).  We overplotted the isochrone 
which best matches the
cluster features (black solid line). }
\label{fig:fig8}
\end{figure}
\end{landscape}

\begin{landscape}
\begin{figure} 
\caption{CMDs and CC diagrams for stars measured in the field of ESO\,437-61. Symbols
are as in Fig.~\ref{fig:fig2}.}
\label{fig:fig9}
\end{figure}
\end{landscape}

\begin{landscape}
\begin{figure} 
\caption{CMDs and CC diagrams for stars measured in the field of ESO\,493-3. Symbols
are as in Fig.~\ref{fig:fig2}. Stars with {\it Gaia} parallaxes and
proper motions are numbered from \#1 to 7. (see text for details).}
\label{fig:fig10}
\end{figure}
\end{landscape}

\begin{figure} 
\caption{CMD for stars measured in the field of ESO\,561-5. Symbols
are as in Fig.~\ref{fig:fig2}.  We overplotted the isochrone which best matches the
cluster features (black solid line). }
\label{fig:fig11}
\end{figure}

\begin{landscape}
\begin{figure} 
\caption{CMDs and CC diagrams for stars measured in the field of Hogg\,9. Symbols
are as in Fig.~\ref{fig:fig2}. Star with {\it Gaia} parallax and
proper motions is numbered \#1. (see text for details).}
\label{fig:fig12}
\end{figure}
\end{landscape}

\begin{landscape}
\begin{figure} 
\caption{CMDs and CC diagrams for stars measured in the field of Miller\,1. Symbols
are as in Fig.~\ref{fig:fig2}.}
\label{fig:fig13}
\end{figure}
\end{landscape}

\begin{landscape}
\begin{figure} 
\caption{CMDs and CC diagrams for stars measured in the field of NGC\,5269. Symbols
are as in Fig.~\ref{fig:fig2}. Stars with {\it Gaia} parallaxes and
proper motions are numbered from \#1 to 4 (see text for details).  We overplotted the 
isochrone which best matches the
cluster features (black solid line). }
\label{fig:fig14}
\end{figure}
\end{landscape}

\begin{landscape}
\begin{figure} 
\caption{CMDs and CC diagrams for stars measured in the field of Ruprecht\,5. Symbols
are as in Fig.~\ref{fig:fig2}.}
\label{fig:fig15}
\end{figure}
\end{landscape}

\begin{landscape}
\begin{figure} 
\caption{CMDs and CC diagrams for stars measured in the field of Ruprecht\,15. Symbols
are as in Fig.~\ref{fig:fig2}. Stars with {\it Gaia} parallaxes and
proper motions is numbered \#1. (see text for details).}
\label{fig:fig16}
\end{figure}
\end{landscape}

%trim .ps CMD & CCD figures 
%85 180 592 718    1
%310 180 592 718   2
%85 440 592 718    3
%310 440 592 718   4

%%%%%%%%%%%%%%%%%%%%%%%%%%%%%%%%%%%%%%%%%%%%%%%%%%

%%%%%%%%%%%%%%%%% APPENDICES %%%%%%%%%%%%%%%%%%%%%


\appendix

%%%%%%%%%%%%%%%%%%%%%%%%%%%%%%%%%%%%%%%%%%%%%%%%%%

% Don't change these lines
\bsp	% typesetting comment
\label{lastpage}

% End of mnras_template.tex

\end{document}